\documentclass[pra,nopacs,twocolumn,nofootinbib]{revtex4}
\usepackage{graphicx}
\usepackage{bm,amsfonts,amsmath,amssymb}
\usepackage{dcolumn}
\usepackage{natbib}

\usepackage
{hyperref}
\hypersetup{
    colorlinks=true,
    linkcolor=blue,
    citecolor=red,
    filecolor=magenta,
    urlcolor=magenta,
}
\newcommand{\Eref}[1]{Eq.~(\ref{#1})}
\newcommand{\Tref}[1]{Table~\ref{#1}}

\def\vkapp{\varkappa}

\begin{document}
\title{Calculation of francium hyperfine anomaly}

\author{E. A. Konovalova$^1$}
\author{Yu. A. Demidov$^{1,2}$}
\author{M. G. Kozlov$^{1,2}$}
\author{A. E. Barzakh$^1$}

\affiliation{$^1$ Petersburg Nuclear Physics Institute of NRC ``Kurchatov center'', Gatchina, Leningrad District 188300, Russia}
\affiliation{$^2$ St.~Petersburg Electrotechnical University ``LETI'', Prof. Popov Str. 5, 197376 St.~Petersburg}

\date{\today}

\begin{abstract}
The Dirac-Hartree-Fock plus many-body perturbation theory (DHF+MBPT) method has been used to calculate hyperfine structure constants for Fr. 
Calculated hyperfine structure anomaly for hydrogen-like ion has been shown to be in good agreement with analytical expressions.
It has been shown that the ratio of the anomalies for $s$ and $p_{1/2}$ states is weakly dependent on the principal quantum number.
Finally, we estimate Bohr--Weisskopf corrections for several Fr isotopes.
Our results may be used to improve experimental accuracy for the nuclear $g$ factors of short-lived isotopes.
\end{abstract}

\maketitle

\section{Introduction}
In recent years, the precision achieved in laser spectroscopy experiments coupled with 
advances in atomic theory has enabled new atomic physics based tests of nuclear models. 
The hyperfine structure constants and isotope shifts are highly sensitive to the changes of charge and magnetization distributions inside the nucleus
because they depend on the behavior of the electron wave function in this region.
The hyperfine structure (HFS) measurements can serve as very useful tool for 
understanding of shape coexistence phenomena in atomic nuclei~\cite{shape_coh}.

The ratio of magnetic hyperfine constants $A$ for different isotopes is usually assumed to be equal to the ratio of their nuclear $g$ factors $g_I = \mu/I$, 
where $\mu$ and $I$ are magnetic moment and spin of the nucleus.
However, this is true only for the point-like nucleus. 
For the finite nucleus one should take into account (i) distribution of the magnetization inside the nucleus and 
(ii) dependence of the electron wave function on the nuclear charge radius. 
Former correction is called magnetic, or Bohr--Weisskopf (BW) correction~\cite{BW50} and the latter one is called charge, 
or Breit-Rosenthal (BR) correction~\cite{RB32,CS49}).
These corrections break proportionality between magnetic hyperfine constants and nuclear $g$ factors.
This phenomenon is called hyperfine anomaly (HFA)~\cite{BW50}. 
Below we discuss how to calculate HFA for many-electron atoms with 
available atomic package~\cite{KPST15}, which is based on the original Dirac-Hartree-Fock 
code~\cite{BDT77}. This package was often used to calculate different atomic properties including HFS
constants of Tl~\cite{DFKP98,KPJ01}, Yb~\cite{PRK99a}, Mg~\cite{KPWAT15}, and Pb~\cite{PKST16}.
 
We study francium atom, because 
there are comprehensive experimental data~\cite{fr_anom,fr_anom2,fr_laser_spectr,fr_laser_spectr2,fr_exp,lu1997} and many theoretical calculations~\cite{fr210,sahoo15,pendrill_fr,dzuba84}
for this isotopic chain.
In particular, changes of the nuclear charge radii in the Fr isotopic series were calculated from isotope shift measurements~\cite{fr_FS,kalita17} 
and absolute values of the nuclear charge radii were obtained~\cite{fr_R_N}.

\section{ Theory and methods}

It is generally accepted that the observed HFS constant $A$ can be written in the following form:
 \begin{align}
 \label{hfs_Shabaev_1}
 A = g_I {\cal A}_0 (1-\delta) (1-\epsilon).
 \end{align}
Here $g_I$ is a nuclear $g$ factor, $g_I{\cal A}_0$ is a HFS constant for the point-like nucleus, 
$\delta$ and $\epsilon$ are the nuclear charge distribution (BR) and magnetization distribution (BW) corrections respectively.
${\cal A}_0$ is independent on the nuclear $g$ factor.
In the case of hydrogen-like ions the expression for ${\cal A}_0$ was obtained in the analytical form by Shabaev~\cite{Sha94}:
 \begin{align}
 \label{hfs_Shabaev_2}
 {\cal A}_0 = \frac{\alpha (\alpha Z)^3}{j(j+1)}\,
 \frac{m}{m_p}\,
 \frac{\vkapp(2\vkapp(\gamma +n_r) - N)}
 {N^4\gamma(4\gamma^2 -1)}\, mc^2.
 \end{align}
Here $\alpha$ is the fine-structure constant, $Z$ is the nuclear charge, $m$ and $m_p$ are electron and proton masses, $j$ is the total electron angular momentum, 
$\vkapp$ is a relativistic quantum number, $N= \sqrt{n_{r}^2 + 2n_r\gamma + \vkapp^2}$, $n_r$ is a radial quantum number, $\gamma = \sqrt{\vkapp^2 - (\alpha Z)^2}$.
We use the model of the homogeneously charged ball of the radius $R = \bigl(\frac{5}{3}\langle r^2\rangle\bigr)^{1/2}$. 
The extended nuclear magnetization leads to a modification of the hyperfine interaction. 
It was shown in Refs.\ \cite{odd_odd_fr,MP95} that the corresponding contribution to the HFS constant may be factorized by ``atomic'' and ``nuclear'' factors.
Following Refs.\ \cite{odd_odd_fr,MP95} the corresponding factor $d_\mathrm{nuc}$ depending on the nuclear spin and configuration, was introduced.
Then corrections $\delta$ and $\epsilon$ for a given $Z$ and electron state can be written as \cite{KKDB17}:
 \begin{align}
 \label{H-scalings}
&\delta(R) = b_N R^{2\gamma -1}, \qquad \epsilon(R, d_\mathrm{nuc}) = b_M d_\mathrm{nuc} R^{2\gamma -1},
 \end{align}
where $b_N$ and $b_M$ are factors, which are independent of the nuclear radius and structure. 

It follows from Eqs.\ \eqref{hfs_Shabaev_1} and \eqref{H-scalings}, that if we calculate HFS constant numerically for different $R$ and $d_\mathrm{nuc}$, 
we should get following dependence on the radius in the first order in $\delta$ and $\epsilon$:
 \begin{align}
 \label{hfs_fit_1}
A(g_I,d_\mathrm{nuc},R)= g_I {\cal A}_0 \left(1 - (b_{N} + b_{M}d_\mathrm{nuc}) R^{2\gamma -1}\right).
 \end{align}
Within the point-like magnetic dipole approximation ($d_\mathrm{nuc}=0$) the
Bohr--Weisskopf correction $\epsilon$ is equal to zero, and the HFS constant can be fitted by the function:
 \begin{align}
 \label{A_PD_fit}
 A (1, 0,R) = {\cal A}_0 \left(1 - b_{N} R^{2\gamma -1}\right).
 \end{align}
On the other hand, for $d_\mathrm{nuc} = 1$ one obtains:
 \begin{align}
 \label{A_UD_fit}
 A (1, 1,R) = {\cal A}_0
 \left(1 - (b_N + b_M) R^{2\gamma -1}\right).
 \end{align}
Let us compare HFS constants for two isotopes with nuclear $g$ factors $g_I^{(1)}$ and $g_I^{(2)}$,
slightly different nuclear radii $R^{(1,2)} = R\pm \mathfrak{r}$, and nuclear factors $d_\mathrm{nuc}^{(1)}=d_\mathrm{nuc}^{(2)}=0$:
 $$\frac{A(g_I^{(1)}, 0, R+\mathfrak{r})}{A(g_I^{(2)}, 0, R-\mathfrak{r})}
 \approx 1 + 2\mathfrak{r}\frac{\partial A(g_I^{(1)}, 0, R)/\partial R}{A(g_I^{(2)}, 0, R)}.$$
Then the part of the HFS anomaly related to the change of the nuclear charge distribution $^1\Delta^2_\mathrm{BR}(R)$ is:
\begin{multline}\label{HFS_anom}
    ^1\Delta^2_\mathrm{BR}(R,\mathfrak{r}) \equiv   
    \frac{g_I^{(2)}A(g_I^{(1)}, 0, R+\mathfrak{r})}
    {g_I^{(1)}A(g_I^{(2)}, 0, R-\mathfrak{r})}-1 
    \approx
 \\
    \approx 2(2\gamma -1) b_N R^{2\gamma -2}\mathfrak{r} .
\end{multline}

Nuclear radii of heavy isotopes are typically very close, then $\mathfrak{r} \ll R$, and anomaly \eqref{HFS_anom} is therefore small. 
For isotopes with the same nuclear factors $d_\mathrm{nuc}$ similar dependence  on the nuclear radii holds for the magnetic part of the HFS anomaly $^1\Delta^2_\mathrm{BW}$. 
However, the nuclear factors may significantly vary from one isotope to another.
In this case we can neglect the radial dependence of the magnetic part of the HFS anomaly and write it as 
$^1\Delta^2_\mathrm{BW}(R,d^{(1)}_\mathrm{nuc},d^{(2)}_\mathrm{nuc})$.
Thus, the HFS anomaly can be divided into two terms related to the nuclear charge 
and magnetization distributions:
 \begin{multline}
 ^1\Delta^2(R,\mathfrak{r},d_\mathrm{nuc}^{(1)},d_\mathrm{nuc}^{(2)}) =
 \\= {}^1\Delta^2_\mathrm{BR}(R,\mathfrak{r})  
 +  {}^1\Delta^2_\mathrm{BW}(R, d_\mathrm{nuc}^{(1)},d_\mathrm{nuc}^{(2)}).
 \end{multline}

In this work we calculate the magnetic hyperfine constants and HFS anomalies for 
low-lying states of Fr atom within the Dirac-Hartree-Fock (DHF) approximation and 
the DHF plus many-body perturbation theory (DHF+MBPT) method. 
The effects of the Breit corrections and spin-polarization of the core are also considered.  

\section{Results and discussion}

\subsection{HFS anomaly for H-like francium ion}

\begin{figure}[htb!]
     \centering
     \includegraphics[height=8.5cm]{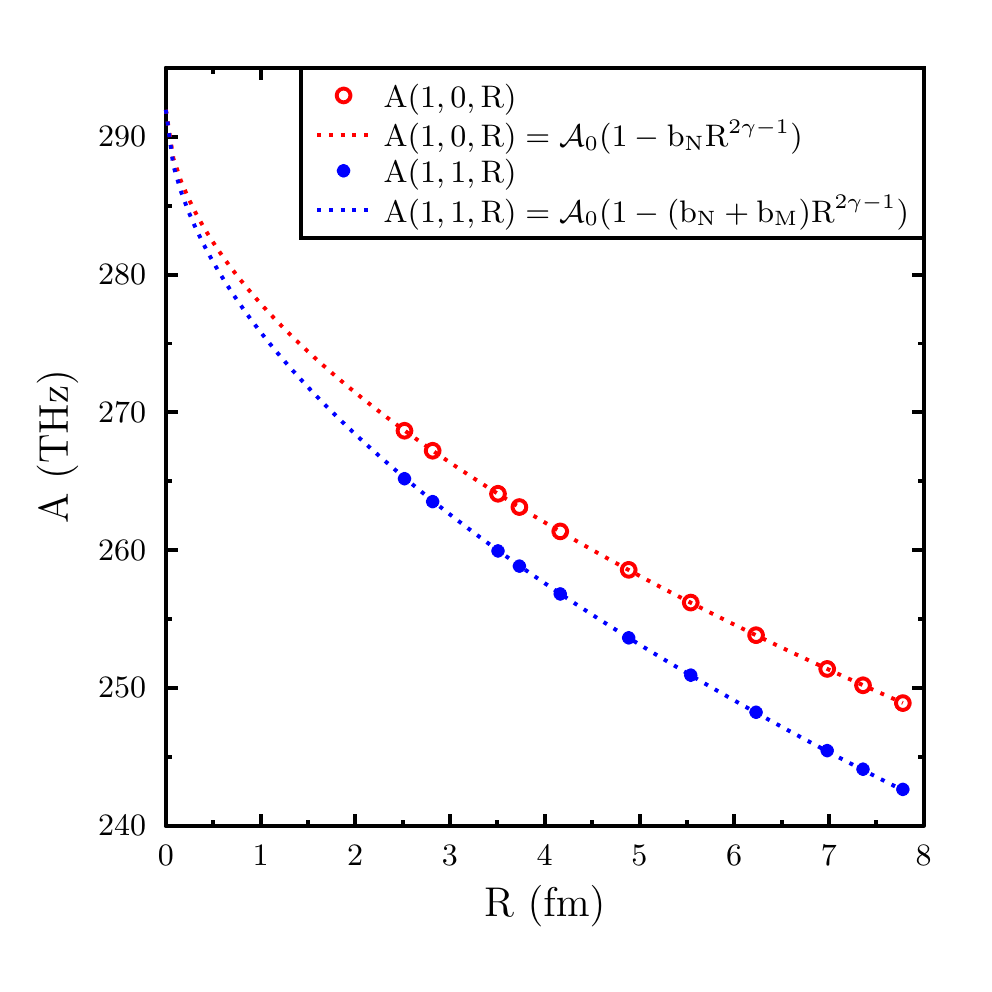}
     \caption{Dependence of the HFS constant $A(g_I, d_\mathrm{nuc}, R)$ for the ground state of H-like Fr ion on the nuclear radius. 
              Dots and circles correspond to the calculated values. Dashed lines correspond to the fits by Eqs.\ (\ref{A_PD_fit}, \ref{A_UD_fit}).
     \label{frg:A_fit}}
\end{figure}

Here we calculate HFS constants of the $1s$, $2s$, and
$2p_{1/2}$ states of Fr$^{86+}$ for the different nuclear radii $R$
and compare our results with analytical expressions from Ref.~\cite{Sha94}. 
Figure \ref{frg:A_fit} shows the dependence of the hyperfine constant $A(1s)$ on the nuclear radius $R$. 
We see very good agreement with Eqs.\ (\ref{A_PD_fit}, \ref{A_UD_fit}).

\begin{table}[!htb]
\caption{\label{tbl:h-like} 
Compilation of the fitting parameters for HFS of the H-like Fr ion: 
${\cal A}_0$ is HFS constant for the point-like nucleus and $g_I = 1$,
$\delta$ and $\epsilon$ are the nuclear charge and magnetization distribution corrections parametrized by coefficients $b_N$ 
and $b_M$. Corrections $\delta$ and $\epsilon$ for $\rm ^{210}Fr^{86+}$ are calculated for $R = 7.1766$ fm~\cite{fr_R_N} and $d_\mathrm{nuc} = 1$.}
\begin{tabular}{lcddd}
\hline \hline
&&\multicolumn{1}{c}{$1s$}
&\multicolumn{1}{c}{$2s$}
&\multicolumn{1}{c}{$2p_{1/2}$}\\
\hline
\\[-3mm]
${\cal A}_0$ (THz)                                                             &fit. & 292.0 & 49.5 & 15.2 \\
                                                                                        &\Eref{hfs_Shabaev_2}& 291.5 & 49.5 & 15.1 \\
$b_{N} \cdot 10^{2}/\mathrm{fm}^{2\gamma-1}$           &fit.& 4.817 & 5.161& 1.650\\
$\rm \delta( ^{210}Fr)$                                                    &fit.& 0.1411 &  0.1512& 0.0483\\
$b_{M} \cdot 10^{2}/\mathrm{fm}^{2\gamma-1}$           &fit.& 0.710 & 0.761& 0.257\\
$\rm \epsilon( ^{210}Fr,d_\mathrm{nuc} = 1)$                 &fit.& 0.0208 & 0.0223 & 0.0075\\
\hline \hline
\end{tabular}
\end{table}

\begin{table*}[tbh]
\caption{\label{tbl_fr} Compilation of the fitting parameters for HFS constants of neutral Fr 
atom: ${\cal A}_0$~(MHz) is HFS constant for point-like nucleus and 
$g_I = 1$; coefficients $b_N$ and $b_M$ $(\mathrm{fm}^{1-2\gamma})$.}
\begin{tabular*}{\textwidth}{@{\extracolsep{\fill}}lccccccccc}
\hline\hline
&\multicolumn{3}{c}{$7s_{1/2}$}&\multicolumn{3}{c}{$7p_{1/2}$}&\multicolumn{3}{c}{$7p_{3/2}$} \\
&${\cal A}_0$&$b_N \cdot 10^{2}$&$b_M \cdot 10^{2}$&
 ${\cal A}_0$&$b_N \cdot 10^{2}$&$b_M \cdot 10^{2}$&
 ${\cal A}_0$&$b_N \cdot 10^{2}$&$b_M \cdot 10^{2}$\\
\hline
DHF&
7894.710& 5.3030& 0.7646&
 746.580& 1.8241& 0.2842&
  55.524& 0.0000& 0.0000\\
  DHF+Br&
7887.813& 5.2989& 0.7642&
 740.345& 1.8204& 0.2837&
  55.151& 0.0000& 0.0000\\
  DHF+MBPT&
10602.174& 4.7502& 0.8584&
 1130.031& 1.5661& 0.3223&
   77.870& 0.0000& 0.0000\\
DHF+MBPT+Br&
10581.950& 4.7013& 0.8506&
 1120.865& 1.5461& 0.3160&
   77.437& 0.0000& 0.0000\\
 \hline
DHF+RPA&
8684.144& 5.1092& 0.8008&
 865.034& 1.6205& 0.2606&
  94.984& 1.2620& 0.2769\\
DHF+Br+RPA&
8682.028& 5.1020& 0.8007&
 861.718& 1.6223& 0.2627&
  94.721& 1.2545& 0.2742\\
DHF+MBPT+RPA&
11518.484&4.6067&0.8844&
 1308.388&1.4018&0.2929&
  132.482&1.2535&0.2919\\
DHF+MBPT+Br+RPA&
11507.415&4.5516&0.8738&
 1300.950&1.3879&0.2891&
  131.988&1.2382&0.2843\\
\hline\hline
\end{tabular*}
\end{table*}

Table~\ref{tbl:h-like} summarizes our results for H-like Fr ion.  For all three states we
see good agreement between analytical values of ${\cal A}_0$ from \Eref{hfs_Shabaev_2} and 
the values obtained from the fit of calculated HFS constants for finite nuclei.
According to our calculations the ratios of the parameters $b_N$ and $b_M$ for $1s$ and $2s$ states
are close to unity: $\frac{b_N(1s)}{b_N(2s)} = 0.933$ and $\frac{b_M
(1s)}{b_M (2s)} = 0.933$. 
This is expected, as in the first approximation the wave functions of the
same symmetry should be proportional to each other inside the
nucleus. Similar ratios for $2s$ and $2p_{1/2}$ states are
$\frac{b_N (2s)}{b_N (2p_{1/2})} = 3.128$ and $\frac{b_M (2s)}{b_M(2p_{1/2})} = 2.961$. 
Again, one can expect that these ratios weakly depend on the principle quantum number.

\subsection{HFS anomaly of neutral francium atom}
The ground configuration of the neutral francium atom is $\rm[Rn]$$7s$.
If we treat francium as an one-electron system with the frozen core, we can do calculation
using Dirac--Hartree--Fock (DHF) method. In this case the dependence of the HFS constants on the nuclear
radius is similar to the one-electron ion.

In the DHF approximation the HFS constant $ A(7p_{3/2}) =  0.56$ GHz is 
very small and practically does not depend on $R$ (see Table \ref{tbl_fr}). 
At the same time, the HFS constants $A(7s)$ and $A(7p_{1/2})$ are well described by Eqs.\ \eqref{A_PD_fit} and \eqref{A_UD_fit}.
According to our calculations, the ratios of coefficients $b_N$ and $b_M$ for $s$ and $p_{1/2}$ waves are close to the respective ratios in H-like ion 
$\frac{b_N(1s)}{b_N(7s)} = 0.908$ and $\frac{b_M(1s)}{b_M(7s)} = 0.929$.
This result is compatible with assertion that the hyperfine anomaly measured for the $s$ states in Rb 
is weakly dependent on the principal quantum number~\cite{GZO07}.
Ratios of the parameters $b_N$ and $b_M$ for $7s$ and $7p_{1/2}$ are: 
$\frac{b_N (7s)}{b_N (7p_{1/2})} = 2.907$ and $\frac{b_M(7s)}{b_M(7p_{1/2})} = 2.690$, while 
for the H-like ion we had 3.128 and 2.961 respectively. 

Situation changes when we include spin-polarization of the core via random phase approximation (RPA) corrections. 
These corrections lead to effective mixing of different partial waves, thus $A(7p_{3/2})$ constant acquires
contributions from the $s$ and $p_{1/2}$ waves. Due to the RPA corrections the value of the constant $A(7p_{3/2})$ is 
significantly changed. At the same time this constant becomes sensitive to the distributions inside the nucleus.
To account for that, we can use Eq.~\eqref{H-scalings} with the same $\gamma$ as for $s$ and $p_{1/2}$ states. 
The RPA corrections for the $7s$ and $7p_{1/2}$ states are smaller than
for $7p_{3/2}$, but they are also significant. 
Due to the RPA corrections the ratios of the parameters $b_N$ and $b_M$ for $7s$ and $7p_{1/2}$ states change by $\sim15\%$:
$\frac{b_N(7s)}{b_N(7p_{1/2})} = 3.153$ and $\frac{b_M(7s)}{b_M(7p_{1/2})} = 3.073$.
 
Core-valence and core-core electron correlations were taken into consederation within DHF+MBPT method~\cite{KPST15}.
Electron correlation corrections significantly change ${\cal A}_0$ values.
The parameters $b_N$ and $b_M$ also change, but ratios of these parameters for the $7s$ and $7p_{1/2}$ states remain stable.
Without RPA corrections these ratios are equal to: $\frac{b_N(7s)}{b_N(7p_{1/2})} = 3.033$ and $\frac{b_M(7s)}{b_M(7p_{1/2})} = 2.663$.
Final ratios were obtained in terms of DHF+MBPT approximation with RPA and Breit corrections:
\begin{align}
 \label{fin_ratio}
 \frac{b_N(7s)}{b_N(7p_{1/2})} = 3.280\,,
 \qquad
 \frac{b_M(7s)}{b_M(7p_{1/2})} = 3.023\,.
 \end{align}
According to ~\citet{pendrill_fr} the ratio of $b_N$ parameters obtained by scaling the Breit-Rosenthal corrections for Tl is equal to 3.2
in good agreement with our result. For the ratio of the $b_M$ parameters the value of 3.0 was used in~\cite{pendrill_fr} also in agreement with our results.

\begin{figure}[tbh]
	     \centering
	     \includegraphics[width=8.5cm]{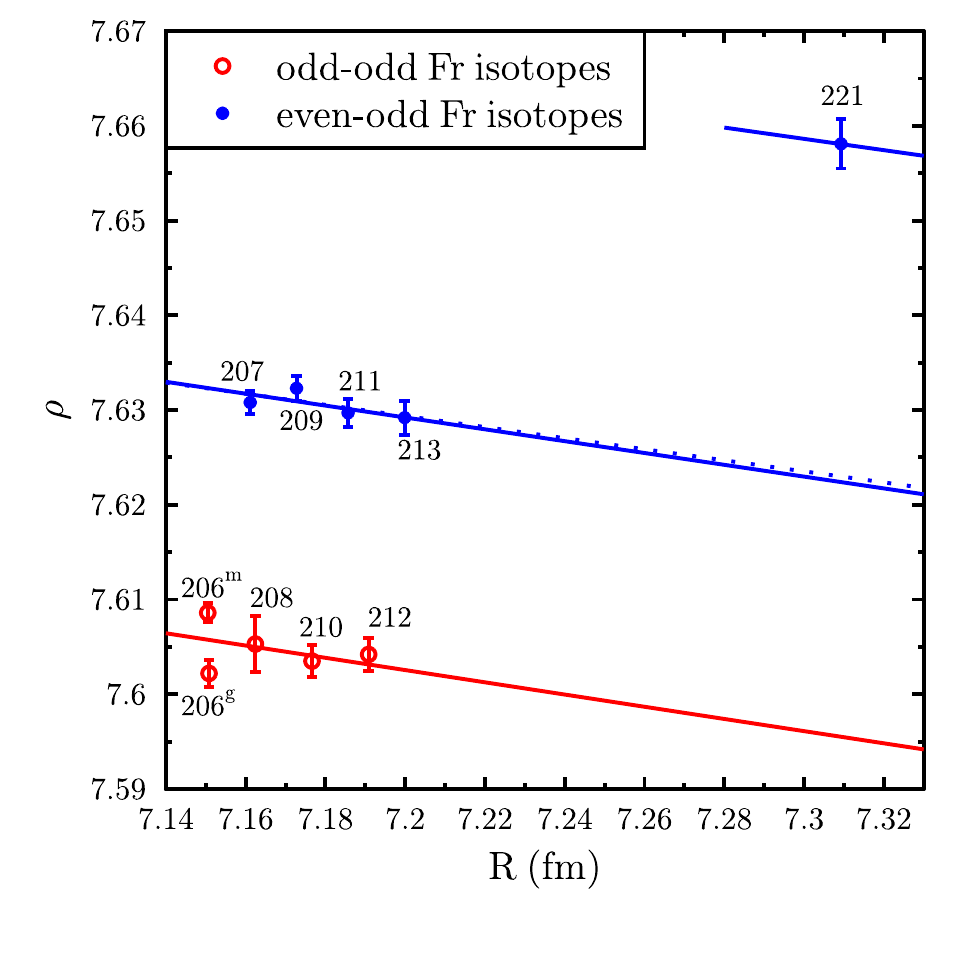}
	     \caption{Experimentally measured ratios $\rho = A(7s)/ A(7p_{1/2})$ for even-odd and odd-odd Fr isotopes~\cite{fr_anom}. 
             The nuclear radii $R$ are taken from Ref.~\cite{fr_R_N}. 
             Lines are the one-parameter fits by \Eref{rho}, dashed line corresponds to two-parameter fit.  
             For even-odd isotopes we use $d_\mathrm{nuc} = 0.3$~\cite{pendrill_fr} and parameters $\rho_0$ (one-parameter fit), 
	     or $\rho_0$ and ($b_N+d_\mathrm{nuc}\,b_M$) (two-parameter fit). 
	     Then for odd-odd isotopes we fix the obtained by the one-parameter fit $\rho_0$ value and fit nuclear factor $d_\mathrm{nuc}$, 
	     with the result $d_\mathrm{nuc} = 0.49$. For $\rm ^{221}Fr$ the fit gives $d_\mathrm{nuc} = 0.05$.
\label{fgr:rho}}
 \end{figure}

Information about parameters $b_N$ and $b_M$ can be extracted from the experimentally measured 
ratio of HFS constants $\rho = {A(7s)}/{A(7p_{1/2})}$.
This ratio can be written as a function of nuclear radius and nuclear factor:
\begin{multline}\label{rho}
	1-\frac{\rho(R, d_\mathrm{nuc})}{\rho_0}  
	\approx  \big(b_N(7s) - b_N(7p_{1/2})\big)R^{2\gamma -1}+\\
        +d_\mathrm{nuc}\big( b_M(7s)-b_M(7p_{1/2})\big) R^{2\gamma -1},
\end{multline}
where $\rho_0 = {\cal A}_0 (7s)/{\cal A}_0 (7p_{1/2})$.
Several experimentally measured values of $\rho$ for odd-odd and even-odd isotopes 
\cite{fr_anom} and corresponding fits by \Eref{rho} are presented in Fig.\ \ref{fgr:rho}. 

Even-odd Fr isotopes with neutron number $N \le 126 \, (A\le 213)$ have spin $I = 9/2$ and nearly constant magnetic moments. When going from $A = 213$ to $A = 207$, the magnetic moment $\mu(A, 9/2)$ changes only by 3\%. 
Their ground states are regarded as a pure shell-model $h_{9/2}$ states, therefore one can assume that $d_\mathrm{nuc}$ factor is also constant within 3\% limits. Factor $d_\mathrm{nuc}$ was calculated by the simple shell-model formula~\cite{pendrill_fr}: $d_\mathrm{nuc}$ = 0.3 for $A = 207-213$. Then the one-parameter fit with $\rho_0$ as free parameter gives us the following relation:
 $\rho = 8.456\,(1-0.033\,R^{2\gamma-1})$, 
where we used our final results for $b_N$ and $b_M$ from \Tref{tbl_fr}, or $\rho = 8.404\,(1-0.031\,R^{2\gamma-1})$ within two-parameter fit. Comparing these two results we can estimate the error bars for fitting parameters to be:  $\rho_0 = 8.43(3)$ and $b_N+d_\mathrm{nuc}\,b_M = 0.032(1)$. Note that the theoretical value of $\rho_0$ obtained within DHF+MBPT+Br+RPA method is equal to 8.85, which is 5\% larger.
Taking into account the possible change of the $d_\mathrm{nuc}$ factor (~3\%) and its possible deviation from the shell-model value,
the correspondence between fitted and calculated $\rho$ values should be regarded as satisfactorily.

We used formulas from  Ref.\ \cite{odd_odd_fr} to calculate $d_\mathrm{nuc}$ factor for the odd-odd Fr isotopes. 
Spins and configurations for odd-odd Fr isotopes with $A = 206-212$ are different ($I$ = 5, 6, 7, 3). Correspondingly, $d_\mathrm{nuc}$ factor is different for different isotopes. 
However, it can be shown that for all these cases $d_\mathrm{nuc} = 0.5(1)$. 
To check the general applicability of our approach we used the same nuclear factor for the all considered odd-odd Fr isotopes.
We fix $\rho_0$ obtained for even-odd Fr isotopes and fit nuclear factor for odd-odd ones which gives us $d_\mathrm{nuc} = 0.49$ in agreement with the shell-model estimation.
The deviation of the experimental $\rho$ values for $\rm ^{206m}Fr$ and $\rm ^{206g}Fr$ from the fit line (see Fig.\ \ref{fgr:rho}) is obviously connected 
with the structural changes in these nuclei resulting in the changes of the $d_\mathrm{nuc}$ factor (see discussion in Ref.\ \cite{fr_anom}).
For $\rm ^{221}Fr$ the fit gives $d_\mathrm{nuc} = 0.05$. 
This result can be of a particular interest for nuclear physics and more detailed analysis will be presented in the forthcoming paper. 

\begin{table}[tbh]
\caption{\label{tbl:fr210}
Calculated HFS constants for low-lying states of neutral $\rm ^{210}Fr$. We use a point-like magnetic dipole approximation,
as described by Eq.~\eqref{A_PD_fit}, and assume that $R=7.1766$~fm and $g_I = 0.733$.
Then, in the final results we add Bohr--Weisskopf correction for $d_\mathrm{nuc}=0.49$.
Available experimental data and other theoretical relativistic coupled-cluster results are also presented.
}
\begin{tabular}{lccc}
\hline \hline
Method & $A(7s)/g_I$&$A(7p_{1/2})/g_I$&$A(7p_{3/2})/g_I$\\
       &(MHz)&(MHz)&(MHz) \\
\hline
DHF             &6668.56& 706.70& 55.52\\
DHF+Br          &6659.37& 700.78& 55.15\\
DHF+MBPT        &9127.20&1078.20& 77.87\\
DHF+MBPT+Br     &9124.92&1070.11& 77.44\\
DHF+RPA         &7491.91& 837.58& 95.43\\
DHF+Br+RPA      &7496.66& 833.22& 94.73\\
DHF+MBPT+RPA    &9964.42&1254.67&127.62\\
DHF+MBPT+Br+RPA &9973.44&1248.07&127.20\\
FINAL (BR \& BW)           &9849.57&1242.94&126.69\\
\hline
Theory$^*$~\cite{fr210}  &9927 &-- & -- \\
Theory~\cite{sahoo15}&  9885.24 & 1279.56  &  104.28 \\
\hline
Experiment~\cite{coc85,fr_anom2,fr_R_N} & 9856(113) & 1296(15) & 106.8(13) \\
\hline \hline
\end{tabular}
$^{*}$ In this Ref.\ the charge and magnetization distributions were modeled by the same Fermi distribution.
\end{table}

The accuracy reached in our calculations of HFS constants for neutral Fr can be 
estimated in comparison with  available experimental and theoretical data presented in 
Table~\ref{tbl:fr210}. 
Due to Bohr--Weisskopf correction calculated $A(7s)$  and $A(7p_{1/2})$ constants of $\rm ^{210}Fr$ are reduced approximately by 1.24\% and 0.41\% respectively. 
Thus, within the DHF+MBPT+Br+RPA method we obtain following final values: $A(7s)/g_I=9849.57$~MHz and $A(7p_{1/2})/g_I=1242.94$~MHz. 
We see that the accuracy of the theory for the $7s$ state is an order of magnitude higher, than for the $7p_{1/2}$ state.

\section{Conclusions}
In this work we use the method developed in Ref.\ \cite{KKDB17} to calculate the hyperfine anomaly by the analysis of the HFS
constants of Fr as functions of nuclear radius. The HFA in this method 
can be parametrized by coefficients $b_N$ and $b_M$. We tested our method by calculating HFS constants 
of H-like francium ion and obtained fairly good agreement with analytical expression from
Ref.\ \cite{Sha94}. Then we made calculations for neutral Fr, described as a system with one valence electron.
We show that the ratios of $b_N(7s)/b_N(7p_{1/2})$ and $b_M(7s)/b_M(7p_{1/2})$ are practically the same,
as in H-like ion and rather stable within the DHF and DHF+MBPT approximations. 
However when we include spin-polarization of the core by means of RPA corrections, these ratios change by 10 -- 15\%.

The corrections caused by redistribution of the magnetization inside the nucleus were estimated 
using experimentally measured ratio of the HFS constants $A(7s)/A(7p_{1/2})$.
Estimated Bohr--Weisskopf corrections for odd-odd francium isotopes 206, 208, 210, and 212 were 
found to be 1.62 times larger, then for even-odd isotopes 207, 209, 211, and 213. The Bohr--Weisskopf 
correction for $\rm ^{221}Fr$ is significantly smaller, than for other even-odd isotopes. 
This information can be used to obtain more accurate values for the nuclear $g$ factors of the Fr isotopes from the ratios of the HFS constants.
The reliability of the applied method enables one to determine the nuclear factor $d_\mathrm{nuc}$ which gives important nuclear-structure information 
and may be compared with the theoretical predictions.
\acknowledgments
Thanks are due to Vladimir Shabaev, Ilya Tupitsyn and Leonid Skripnikov for helpful discussions. 
The work was supported by the Foundation for the advancement of theoretical physics ``BASIS''  (grant \# 17-11-136).


\begin{thebibliography}{30}
\expandafter\ifx\csname natexlab\endcsname\relax\def\natexlab#1{#1}\fi
\expandafter\ifx\csname bibnamefont\endcsname\relax
  \def\bibnamefont#1{#1}\fi
\expandafter\ifx\csname bibfnamefont\endcsname\relax
  \def\bibfnamefont#1{#1}\fi
\expandafter\ifx\csname citenamefont\endcsname\relax
  \def\citenamefont#1{#1}\fi
\expandafter\ifx\csname url\endcsname\relax
  \def\url#1{\texttt{#1}}\fi
\expandafter\ifx\csname urlprefix\endcsname\relax\def\urlprefix{URL }\fi
\providecommand{\bibinfo}[2]{#2}
\providecommand{\eprint}[2][]{\url{#2}}

\bibitem[{\citenamefont{Andreyev et~al.}(2000)\citenamefont{Andreyev, Huyse,
  Van~Duppen, Weissman, Ackermann, Gerl, Hessberger, Hofmann, Kleinb{\"o}hl,
  M{\"u}nzenberg et~al.}}]{shape_coh}
\bibinfo{author}{\bibfnamefont{A.}~\bibnamefont{Andreyev}},
  \bibinfo{author}{\bibfnamefont{M.}~\bibnamefont{Huyse}},
  \bibinfo{author}{\bibfnamefont{P.}~\bibnamefont{Van~Duppen}},
  \bibinfo{author}{\bibfnamefont{L.}~\bibnamefont{Weissman}},
  \bibinfo{author}{\bibfnamefont{D.}~\bibnamefont{Ackermann}},
  \bibinfo{author}{\bibfnamefont{J.}~\bibnamefont{Gerl}},
  \bibinfo{author}{\bibfnamefont{F.}~\bibnamefont{Hessberger}},
  \bibinfo{author}{\bibfnamefont{S.}~\bibnamefont{Hofmann}},
  \bibinfo{author}{\bibfnamefont{A.}~\bibnamefont{Kleinb{\"o}hl}},
  \bibinfo{author}{\bibfnamefont{G.}~\bibnamefont{M{\"u}nzenberg}},
  \bibnamefont{et~al.}, \bibinfo{journal}{Nature}
  \textbf{\bibinfo{volume}{405}}, \bibinfo{pages}{430} (\bibinfo{year}{2000}).

\bibitem[{\citenamefont{Bohr and Weisskopf}(1950)}]{BW50}
\bibinfo{author}{\bibfnamefont{A.}~\bibnamefont{Bohr}} \bibnamefont{and}
  \bibinfo{author}{\bibfnamefont{V.~F.} \bibnamefont{Weisskopf}},
  \bibinfo{journal}{Phys. Rev.} \textbf{\bibinfo{volume}{77}},
  \bibinfo{pages}{94} (\bibinfo{year}{1950}).

\bibitem[{\citenamefont{Rosenthal and Breit}(1932)}]{RB32}
\bibinfo{author}{\bibfnamefont{J.~E.} \bibnamefont{Rosenthal}}
  \bibnamefont{and} \bibinfo{author}{\bibfnamefont{G.}~\bibnamefont{Breit}},
  \bibinfo{journal}{Phys. Rev.} \textbf{\bibinfo{volume}{41}},
  \bibinfo{pages}{459} (\bibinfo{year}{1932}).

\bibitem[{\citenamefont{Crawford and Schawlow}(1949)}]{CS49}
\bibinfo{author}{\bibfnamefont{M.}~\bibnamefont{Crawford}} \bibnamefont{and}
  \bibinfo{author}{\bibfnamefont{A.}~\bibnamefont{Schawlow}},
  \bibinfo{journal}{Phys. Rev.} \textbf{\bibinfo{volume}{76}},
  \bibinfo{pages}{1310} (\bibinfo{year}{1949}).

\bibitem[{\citenamefont{Shabaev}(1994)}]{Sha94}
\bibinfo{author}{\bibfnamefont{V.~M.} \bibnamefont{Shabaev}},
  \bibinfo{journal}{J. Phys. B} \textbf{\bibinfo{volume}{27}},
  \bibinfo{pages}{5825} (\bibinfo{year}{1994}).

\bibitem[{\citenamefont{Kozlov et~al.}(2015)\citenamefont{Kozlov, Porsev,
  Safronova, and Tupitsyn}}]{KPST15}
\bibinfo{author}{\bibfnamefont{M.}~\bibnamefont{Kozlov}},
  \bibinfo{author}{\bibfnamefont{S.}~\bibnamefont{Porsev}},
  \bibinfo{author}{\bibfnamefont{M.}~\bibnamefont{Safronova}},
  \bibnamefont{and} \bibinfo{author}{\bibfnamefont{I.}~\bibnamefont{Tupitsyn}},
  \bibinfo{journal}{Comput. Phys. Commun.}
  \textbf{\bibinfo{volume}{195}}, \bibinfo{pages}{199} (\bibinfo{year}{2015}),
  ISSN \bibinfo{issn}{0010-4655}.

\bibitem[{\citenamefont{Bratsev et~al.}(1977)\citenamefont{Bratsev, Deyneka,
  and Tupitsyn}}]{BDT77}
\bibinfo{author}{\bibfnamefont{V.~F.} \bibnamefont{Bratsev}},
  \bibinfo{author}{\bibfnamefont{G.~B.} \bibnamefont{Deyneka}},
  \bibnamefont{and} \bibinfo{author}{\bibfnamefont{I.~I.}
  \bibnamefont{Tupitsyn}}, \bibinfo{journal}{Bull. Acad. Sci. USSR, Phys. Ser.}
  \textbf{\bibinfo{volume}{41}}, \bibinfo{pages}{173} (\bibinfo{year}{1977}).

\bibitem[{\citenamefont{Dzuba et~al.}(1998)\citenamefont{Dzuba, Flambaum,
  Kozlov, and Porsev}}]{DFKP98}
\bibinfo{author}{\bibfnamefont{V.~A.} \bibnamefont{Dzuba}},
  \bibinfo{author}{\bibfnamefont{V.~V.} \bibnamefont{Flambaum}},
  \bibinfo{author}{\bibfnamefont{M.~G.} \bibnamefont{Kozlov}},
  \bibnamefont{and} \bibinfo{author}{\bibfnamefont{S.~G.}
  \bibnamefont{Porsev}}, \bibinfo{journal}{Sov. Phys.--JETP}
  \textbf{\bibinfo{volume}{87}}, \bibinfo{pages}{885} (\bibinfo{year}{1998}).

\bibitem[{\citenamefont{Kozlov et~al.}(2001)\citenamefont{Kozlov, Porsev, and
  Johnson}}]{KPJ01}
\bibinfo{author}{\bibfnamefont{M.~G.} \bibnamefont{Kozlov}},
  \bibinfo{author}{\bibfnamefont{S.~G.} \bibnamefont{Porsev}},
  \bibnamefont{and} \bibinfo{author}{\bibfnamefont{W.~R.}
  \bibnamefont{Johnson}}, \bibinfo{journal}{Phys. Rev. A}
  \textbf{\bibinfo{volume}{64}}, \bibinfo{pages}{052107}
  (\bibinfo{year}{2001}), \eprint{arXiv: physics/0105090}.

\bibitem[{\citenamefont{Porsev et~al.}(1999)\citenamefont{Porsev, Rakhlina, and
  Kozlov}}]{PRK99a}
\bibinfo{author}{\bibfnamefont{S.~G.} \bibnamefont{Porsev}},
  \bibinfo{author}{\bibfnamefont{Y.~G.} \bibnamefont{Rakhlina}},
  \bibnamefont{and} \bibinfo{author}{\bibfnamefont{M.~G.}
  \bibnamefont{Kozlov}}, \bibinfo{journal}{J. Phys. B}
  \textbf{\bibinfo{volume}{32}}, \bibinfo{pages}{1113} (\bibinfo{year}{1999}),
  \eprint{arXiv: physics/9810011}.

\bibitem[{\citenamefont{{Kj{\o}ller} et~al.}(2015)\citenamefont{{Kj{\o}ller},
  {Porsev}, {Westergaard}, {Andersen}, and {Thomsen}}}]{KPWAT15}
\bibinfo{author}{\bibfnamefont{N.~K.} \bibnamefont{{Kj{\o}ller}}},
  \bibinfo{author}{\bibfnamefont{S.~G.} \bibnamefont{{Porsev}}},
  \bibinfo{author}{\bibfnamefont{P.~G.} \bibnamefont{{Westergaard}}},
  \bibinfo{author}{\bibfnamefont{N.}~\bibnamefont{{Andersen}}},
  \bibnamefont{and} \bibinfo{author}{\bibfnamefont{J.~W.}
  \bibnamefont{{Thomsen}}}, \bibinfo{journal}{\pra}
  \textbf{\bibinfo{volume}{91}}, \bibinfo{pages}{032515}
  (\bibinfo{year}{2015}).

\bibitem[{\citenamefont{Porsev et~al.}(2016)\citenamefont{Porsev, Kozlov,
  Safronova, and Tupitsyn}}]{PKST16}
\bibinfo{author}{\bibfnamefont{S.~G.} \bibnamefont{Porsev}},
  \bibinfo{author}{\bibfnamefont{M.~G.} \bibnamefont{Kozlov}},
  \bibinfo{author}{\bibfnamefont{M.~S.} \bibnamefont{Safronova}},
  \bibnamefont{and} \bibinfo{author}{\bibfnamefont{I.~I.}
  \bibnamefont{Tupitsyn}}, \bibinfo{journal}{Phys. Rev. A}
  \textbf{\bibinfo{volume}{93}}, \bibinfo{pages}{012501}
  (\bibinfo{year}{2016}), \eprint{1510.06679}.

\bibitem[{\citenamefont{Zhang et~al.}(2015)\citenamefont{Zhang, Tandecki,
  Collister, Aubin, Behr, Gomez, Gwinner, Orozco, Pearson, Sprouse
  et~al.}}]{fr_anom}
\bibinfo{author}{\bibfnamefont{J.}~\bibnamefont{Zhang}},
  \bibinfo{author}{\bibfnamefont{M.}~\bibnamefont{Tandecki}},
  \bibinfo{author}{\bibfnamefont{R.}~\bibnamefont{Collister}},
  \bibinfo{author}{\bibfnamefont{S.}~\bibnamefont{Aubin}},
  \bibinfo{author}{\bibfnamefont{J.}~\bibnamefont{Behr}},
  \bibinfo{author}{\bibfnamefont{E.}~\bibnamefont{Gomez}},
  \bibinfo{author}{\bibfnamefont{G.}~\bibnamefont{Gwinner}},
  \bibinfo{author}{\bibfnamefont{L.}~\bibnamefont{Orozco}},
  \bibinfo{author}{\bibfnamefont{M.}~\bibnamefont{Pearson}},
  \bibinfo{author}{\bibfnamefont{G.}~\bibnamefont{Sprouse}},
  \bibnamefont{et~al.}, \bibinfo{journal}{Phys. Rev. Lett.}
  \textbf{\bibinfo{volume}{115}}, \bibinfo{pages}{042501}
  (\bibinfo{year}{2015}).

\bibitem[{\citenamefont{Grossman et~al.}(1999)\citenamefont{Grossman, Orozco,
  Pearson, Simsarian, Sprouse, and Zhao}}]{fr_anom2}
\bibinfo{author}{\bibfnamefont{J.}~\bibnamefont{Grossman}},
  \bibinfo{author}{\bibfnamefont{L.}~\bibnamefont{Orozco}},
  \bibinfo{author}{\bibfnamefont{M.}~\bibnamefont{Pearson}},
  \bibinfo{author}{\bibfnamefont{J.}~\bibnamefont{Simsarian}},
  \bibinfo{author}{\bibfnamefont{G.}~\bibnamefont{Sprouse}}, \bibnamefont{and}
  \bibinfo{author}{\bibfnamefont{W.}~\bibnamefont{Zhao}},
  \bibinfo{journal}{Phys. Rev. Lett.} \textbf{\bibinfo{volume}{83}},
  \bibinfo{pages}{935} (\bibinfo{year}{1999}).

\bibitem[{\citenamefont{Budin{\v{c}}evi{\'c}
  et~al.}(2014)\citenamefont{Budin{\v{c}}evi{\'c}, Billowes, Bissell, Cocolios,
  De~Groote, De~Schepper, Fedosseev, Flanagan, Franchoo, Ruiz
  et~al.}}]{fr_laser_spectr}
\bibinfo{author}{\bibfnamefont{I.}~\bibnamefont{Budin{\v{c}}evi{\'c}}},
  \bibinfo{author}{\bibfnamefont{J.}~\bibnamefont{Billowes}},
  \bibinfo{author}{\bibfnamefont{M.}~\bibnamefont{Bissell}},
  \bibinfo{author}{\bibfnamefont{T.~E.} \bibnamefont{Cocolios}},
  \bibinfo{author}{\bibfnamefont{R.}~\bibnamefont{De~Groote}},
  \bibinfo{author}{\bibfnamefont{S.}~\bibnamefont{De~Schepper}},
  \bibinfo{author}{\bibfnamefont{V.~N.} \bibnamefont{Fedosseev}},
  \bibinfo{author}{\bibfnamefont{K.~T.} \bibnamefont{Flanagan}},
  \bibinfo{author}{\bibfnamefont{S.}~\bibnamefont{Franchoo}},
  \bibinfo{author}{\bibfnamefont{R.~G.} \bibnamefont{Ruiz}},
  \bibnamefont{et~al.}, \bibinfo{journal}{Phys. Rev. C}
  \textbf{\bibinfo{volume}{90}}, \bibinfo{pages}{014317}
  (\bibinfo{year}{2014}).

\bibitem[{\citenamefont{De~Groote et~al.}(2015)\citenamefont{De~Groote,
  Budin{\v{c}}evi{\'c}, Billowes, Bissell, Cocolios, Farooq-Smith, Fedosseev,
  Flanagan, Franchoo, Ruiz et~al.}}]{fr_laser_spectr2}
\bibinfo{author}{\bibfnamefont{R.}~\bibnamefont{De~Groote}},
  \bibinfo{author}{\bibfnamefont{I.}~\bibnamefont{Budin{\v{c}}evi{\'c}}},
  \bibinfo{author}{\bibfnamefont{J.}~\bibnamefont{Billowes}},
  \bibinfo{author}{\bibfnamefont{M.}~\bibnamefont{Bissell}},
  \bibinfo{author}{\bibfnamefont{T.~E.} \bibnamefont{Cocolios}},
  \bibinfo{author}{\bibfnamefont{G.~J.} \bibnamefont{Farooq-Smith}},
  \bibinfo{author}{\bibfnamefont{V.}~\bibnamefont{Fedosseev}},
  \bibinfo{author}{\bibfnamefont{K.}~\bibnamefont{Flanagan}},
  \bibinfo{author}{\bibfnamefont{S.}~\bibnamefont{Franchoo}},
  \bibinfo{author}{\bibfnamefont{R.~G.} \bibnamefont{Ruiz}},
  \bibnamefont{et~al.}, \bibinfo{journal}{Phys. Rev. Lett.}
  \textbf{\bibinfo{volume}{115}}, \bibinfo{pages}{132501}
  (\bibinfo{year}{2015}).

\bibitem[{\citenamefont{Flanagan et~al.}(2013)\citenamefont{Flanagan, Lynch,
  Billowes, Bissell, Budin{\v{c}}evi{\'c}, Cocolios, De~Groote, De~Schepper,
  Fedosseev, Franchoo et~al.}}]{fr_exp}
\bibinfo{author}{\bibfnamefont{K.}~\bibnamefont{Flanagan}},
  \bibinfo{author}{\bibfnamefont{K.}~\bibnamefont{Lynch}},
  \bibinfo{author}{\bibfnamefont{J.}~\bibnamefont{Billowes}},
  \bibinfo{author}{\bibfnamefont{M.}~\bibnamefont{Bissell}},
  \bibinfo{author}{\bibfnamefont{I.}~\bibnamefont{Budin{\v{c}}evi{\'c}}},
  \bibinfo{author}{\bibfnamefont{T.~E.} \bibnamefont{Cocolios}},
  \bibinfo{author}{\bibfnamefont{R.}~\bibnamefont{De~Groote}},
  \bibinfo{author}{\bibfnamefont{S.}~\bibnamefont{De~Schepper}},
  \bibinfo{author}{\bibfnamefont{V.}~\bibnamefont{Fedosseev}},
  \bibinfo{author}{\bibfnamefont{S.}~\bibnamefont{Franchoo}},
  \bibnamefont{et~al.}, \bibinfo{journal}{Phys. Rev. Lett.}
  \textbf{\bibinfo{volume}{111}}, \bibinfo{pages}{212501}
  (\bibinfo{year}{2013}).

\bibitem[{\citenamefont{Lu et~al.}(1997)\citenamefont{Lu, Corwin, Vogel,
  Wieman, Dinneen, Maddi, and Gould}}]{lu1997}
\bibinfo{author}{\bibfnamefont{Z.-T.} \bibnamefont{Lu}},
  \bibinfo{author}{\bibfnamefont{K.}~\bibnamefont{Corwin}},
  \bibinfo{author}{\bibfnamefont{K.}~\bibnamefont{Vogel}},
  \bibinfo{author}{\bibfnamefont{C.}~\bibnamefont{Wieman}},
  \bibinfo{author}{\bibfnamefont{T.}~\bibnamefont{Dinneen}},
  \bibinfo{author}{\bibfnamefont{J.}~\bibnamefont{Maddi}}, \bibnamefont{and}
  \bibinfo{author}{\bibfnamefont{H.}~\bibnamefont{Gould}},
  \bibinfo{journal}{Phys. Rev. Lett.} \textbf{\bibinfo{volume}{79}},
  \bibinfo{pages}{994} (\bibinfo{year}{1997}).

\bibitem[{\citenamefont{Gomez et~al.}(2008)\citenamefont{Gomez, Aubin, Orozco,
  Sprouse, Iskrenova-Tchoukova, and Safronova}}]{fr210}
\bibinfo{author}{\bibfnamefont{E.}~\bibnamefont{Gomez}},
  \bibinfo{author}{\bibfnamefont{S.}~\bibnamefont{Aubin}},
  \bibinfo{author}{\bibfnamefont{L.}~\bibnamefont{Orozco}},
  \bibinfo{author}{\bibfnamefont{G.}~\bibnamefont{Sprouse}},
  \bibinfo{author}{\bibfnamefont{E.}~\bibnamefont{Iskrenova-Tchoukova}},
  \bibnamefont{and}
  \bibinfo{author}{\bibfnamefont{M.}~\bibnamefont{Safronova}},
  \bibinfo{journal}{Phys. Rev. Lett.} \textbf{\bibinfo{volume}{100}},
  \bibinfo{pages}{172502} (\bibinfo{year}{2008}).

\bibitem[{\citenamefont{Sahoo et~al.}(2015)\citenamefont{Sahoo, Nandy, Das, and
  Sakemi}}]{sahoo15}
\bibinfo{author}{\bibfnamefont{B.}~\bibnamefont{Sahoo}},
  \bibinfo{author}{\bibfnamefont{D.}~\bibnamefont{Nandy}},
  \bibinfo{author}{\bibfnamefont{B.}~\bibnamefont{Das}}, \bibnamefont{and}
  \bibinfo{author}{\bibfnamefont{Y.}~\bibnamefont{Sakemi}},
  \bibinfo{journal}{Phys. Rev. A} \textbf{\bibinfo{volume}{91}},
  \bibinfo{pages}{042507} (\bibinfo{year}{2015}).

\bibitem[{\citenamefont{M{\aa}rtensson-Pendrill}(2000)}]{pendrill_fr}
\bibinfo{author}{\bibfnamefont{A.~M.} \bibnamefont{M{\aa}rtensson-Pendrill}},
  \bibinfo{journal}{Hyperfine Interact.} \textbf{\bibinfo{volume}{127}},
  \bibinfo{pages}{41} (\bibinfo{year}{2000}).

\bibitem[{\citenamefont{Dzuba et~al.}(1984)\citenamefont{Dzuba, Flambaum, and
  Sushkov}}]{dzuba84}
\bibinfo{author}{\bibfnamefont{V.}~\bibnamefont{Dzuba}},
  \bibinfo{author}{\bibfnamefont{V.}~\bibnamefont{Flambaum}}, \bibnamefont{and}
  \bibinfo{author}{\bibfnamefont{O.}~\bibnamefont{Sushkov}},
  \bibinfo{journal}{J. Phys. B}
  \textbf{\bibinfo{volume}{17}}, \bibinfo{pages}{1953} (\bibinfo{year}{1984}).

\bibitem[{\citenamefont{Dzuba et~al.}(2005)\citenamefont{Dzuba, Johnson, and
  Safronova}}]{fr_FS}
\bibinfo{author}{\bibfnamefont{V.}~\bibnamefont{Dzuba}},
  \bibinfo{author}{\bibfnamefont{W.}~\bibnamefont{Johnson}}, \bibnamefont{and}
  \bibinfo{author}{\bibfnamefont{M.}~\bibnamefont{Safronova}},
  \bibinfo{journal}{Phys. Rev. A} \textbf{\bibinfo{volume}{72}},
  \bibinfo{pages}{022503} (\bibinfo{year}{2005}).

\bibitem[{\citenamefont{Kalita et~al.}(2017)\citenamefont{Kalita, Behr,
  Gorelov, Pearson, Dehart, Gwinner, Kossin, Aubin, Gomez, Orozco
  et~al.}}]{kalita17}
\bibinfo{author}{\bibfnamefont{M.}~\bibnamefont{Kalita}},
  \bibinfo{author}{\bibfnamefont{J.}~\bibnamefont{Behr}},
  \bibinfo{author}{\bibfnamefont{A.}~\bibnamefont{Gorelov}},
  \bibinfo{author}{\bibfnamefont{M.}~\bibnamefont{Pearson}},
  \bibinfo{author}{\bibfnamefont{A.}~\bibnamefont{Dehart}},
  \bibinfo{author}{\bibfnamefont{G.}~\bibnamefont{Gwinner}},
  \bibinfo{author}{\bibfnamefont{M.}~\bibnamefont{Kossin}},
  \bibinfo{author}{\bibfnamefont{S.}~\bibnamefont{Aubin}},
  \bibinfo{author}{\bibfnamefont{E.}~\bibnamefont{Gomez}},
  \bibinfo{author}{\bibfnamefont{L.~A.} \bibnamefont{Orozco}},
  \bibnamefont{et~al.}, in \emph{\bibinfo{booktitle}{APS Division of Atomic,
  Molecular and Optical Physics Meeting Abstracts}} (\bibinfo{year}{2017}).

\bibitem[{\citenamefont{Voss et~al.}(2015)\citenamefont{Voss, Buchinger, Cheal,
  Crawford, Dilling, Kortelainen, Kwiatkowski, Leary, Levy, Mooshammer
  et~al.}}]{fr_R_N}
\bibinfo{author}{\bibfnamefont{A.}~\bibnamefont{Voss}},
  \bibinfo{author}{\bibfnamefont{F.}~\bibnamefont{Buchinger}},
  \bibinfo{author}{\bibfnamefont{B.}~\bibnamefont{Cheal}},
  \bibinfo{author}{\bibfnamefont{J.}~\bibnamefont{Crawford}},
  \bibinfo{author}{\bibfnamefont{J.}~\bibnamefont{Dilling}},
  \bibinfo{author}{\bibfnamefont{M.}~\bibnamefont{Kortelainen}},
  \bibinfo{author}{\bibfnamefont{A.}~\bibnamefont{Kwiatkowski}},
  \bibinfo{author}{\bibfnamefont{A.}~\bibnamefont{Leary}},
  \bibinfo{author}{\bibfnamefont{C.}~\bibnamefont{Levy}},
  \bibinfo{author}{\bibfnamefont{F.}~\bibnamefont{Mooshammer}},
  \bibnamefont{et~al.}, \bibinfo{journal}{Phys. Rev. C}
  \textbf{\bibinfo{volume}{91}}, \bibinfo{pages}{044307}
  (\bibinfo{year}{2015}).

\bibitem[{\citenamefont{B{\"u}ttgenbach}(1984)}]{odd_odd_fr}
\bibinfo{author}{\bibfnamefont{S.}~\bibnamefont{B{\"u}ttgenbach}},
  \bibinfo{journal}{Hyperfine Interact.} \textbf{\bibinfo{volume}{20}},
  \bibinfo{pages}{1} (\bibinfo{year}{1984}).

\bibitem[{\citenamefont{M{\aa}rtesson-Pendrill}(1995)}]{MP95}
\bibinfo{author}{\bibfnamefont{A.-M.} \bibnamefont{M{\aa}rtesson-Pendrill}},
  \bibinfo{journal}{Phys. Rev. Lett.} \textbf{\bibinfo{volume}{74}},
  \bibinfo{pages}{2184} (\bibinfo{year}{1995}).

\bibitem[{\citenamefont{Konovalova et~al.}(2017)\citenamefont{Konovalova,
  Kozlov, Demidov, and Barzakh}}]{KKDB17}
\bibinfo{author}{\bibfnamefont{E.}~\bibnamefont{Konovalova}},
  \bibinfo{author}{\bibfnamefont{M.}~\bibnamefont{Kozlov}},
  \bibinfo{author}{\bibfnamefont{Y.}~\bibnamefont{Demidov}}, \bibnamefont{and}
  \bibinfo{author}{\bibfnamefont{A.}~\bibnamefont{Barzakh}},
  \bibinfo{journal}{Rad. Applic.} \textbf{\bibinfo{volume}{2}},
  \bibinfo{pages}{181} (\bibinfo{year}{2017}), ISSN \bibinfo{issn}{2466-4294},
  \eprint{arXiv:1703.10048}.

\bibitem[{\citenamefont{Galv{\'a}n et~al.}(2007)\citenamefont{Galv{\'a}n, Zhao,
  Orozco, G{\'o}mez, Lange, Baumer, and Sprouse}}]{GZO07}
\bibinfo{author}{\bibfnamefont{A.~P.} \bibnamefont{Galv{\'a}n}},
  \bibinfo{author}{\bibfnamefont{Y.}~\bibnamefont{Zhao}},
  \bibinfo{author}{\bibfnamefont{L.}~\bibnamefont{Orozco}},
  \bibinfo{author}{\bibfnamefont{E.}~\bibnamefont{G{\'o}mez}},
  \bibinfo{author}{\bibfnamefont{A.}~\bibnamefont{Lange}},
  \bibinfo{author}{\bibfnamefont{F.}~\bibnamefont{Baumer}}, \bibnamefont{and}
  \bibinfo{author}{\bibfnamefont{G.}~\bibnamefont{Sprouse}},
  \bibinfo{journal}{Phys. Lett. B} \textbf{\bibinfo{volume}{655}},
  \bibinfo{pages}{114} (\bibinfo{year}{2007}).

\bibitem[{\citenamefont{Coc et~al.}(1985)\citenamefont{Coc, Thibault, Touchard,
  Duong, Juncar, Liberman, Pinard, Lerm{\'e}, Vialle, B{\"u}ttgenbach
  et~al.}}]{coc85}
\bibinfo{author}{\bibfnamefont{A.}~\bibnamefont{Coc}},
  \bibinfo{author}{\bibfnamefont{C.}~\bibnamefont{Thibault}},
  \bibinfo{author}{\bibfnamefont{F.}~\bibnamefont{Touchard}},
  \bibinfo{author}{\bibfnamefont{H.}~\bibnamefont{Duong}},
  \bibinfo{author}{\bibfnamefont{P.}~\bibnamefont{Juncar}},
  \bibinfo{author}{\bibfnamefont{S.}~\bibnamefont{Liberman}},
  \bibinfo{author}{\bibfnamefont{J.}~\bibnamefont{Pinard}},
  \bibinfo{author}{\bibfnamefont{J.}~\bibnamefont{Lerm{\'e}}},
  \bibinfo{author}{\bibfnamefont{J.}~\bibnamefont{Vialle}},
  \bibinfo{author}{\bibfnamefont{S.}~\bibnamefont{B{\"u}ttgenbach}},
  \bibnamefont{et~al.}, \bibinfo{journal}{Phys. Lett. B}
  \textbf{\bibinfo{volume}{163}}, \bibinfo{pages}{66} (\bibinfo{year}{1985}).

\end{thebibliography}
\end{document}